# Thermal tuning of the carrier density in Dirac semimetal $Cd_3As_2$ nanoplates

Min Wu[1,2*], Guolin Zheng[1,3*], Zheng Chen[1], Yequn Liu[4], Wenshuai Gao[5], Hongwei Zhang[1], Yuyan Han[1], Lan Wang[3], Jianhui Zhou[1§], Wei Ning[1§], and Mingliang Tian[1,5,6§]

[1]*Anhui Province Key Laboratory of Condensed Matter Physics at Extreme Conditions, High Magnetic Field Laboratory, the Chinese Academy of Science (CAS), Hefei 230031, Anhui, China.*

[2] *Department of physics, University of Science and Technology of China. Hefei 230026, China.*

[3] *School of Science, RMIT university, Melbourne, VIC 3001, Australia.*

[4] *Analytical Instrumentation Center, State Key Laboratory of Coal Conversion, Institute of Coal Chemistry, Chinese Academy of Sciences, Taiyuan 030001, China.*

[5] *Department of Physics, School of Physics and Materials Science, Anhui University, Hefei 230601, Anhui, China.*

[6] *Collaborative Innovation Center of Advanced Microstructures, Nanjing University, Nanjing 210093, China.*

[*] Those authors contribute equally to this work.

[§]To whom correspondence should be addressed. E-mail: jhzhou@hmfl.ac.cn (J.Z.), ningwei@hmfl.ac.cn (W.N.), tianml@hmfl.ac.cn (M.T.).

# Abstract

The tunable carrier density plays a key role in the investigation of novel transport properties in three-dimensional topological semimetals. Here we demonstrate that the carrier density as well as the mobility of Dirac semimetal $Cd_3As_2$ nanoplates can be effectively tuned by the *in-situ* thermal treatment at 350 K for one hour, both showing a non-monotonic evolution with the thermal cycling treatments. The upward shift of Fermi level relative to the Dirac nodes blurs the surface Fermi-arc states, accompanying with an anomalous phase shift of oscillations of bulk states due to the change of topology of electrons. Meanwhile, the peaks of oscillations of bulk longitudinal magnetoresistivity shift at high fields due to their coupling to the oscillations of the surface Fermi-arc states. Our work provides a thermal control knob for manipulations of the quantum states through the carrier density in Dirac semimetal $Cd_3As_2$ at high temperature.

Three dimensional (3D) Dirac semimetal, Cadmium arsenide ($Cd_3As_2$),[1,2,3] has only one pair of Dirac nodes in momentum space with a nearly sphere-like Fermi surface and provides us with an ideal platform to investigate the novel transport properties in 3D Dirac/Weyl semimetals. Due to the nontrivial topology of Dirac/Weyl nodes in momentum space, 3D topological semimetals exhibit a number of unusual physical properties,[4,5,6] some of which have been observed in a series of experiments, such as ultra-high carrier mobility,[7-10] negative longitudinal magnetoresistivity (LMR),[11-18] planar Hall effect,[18,19] and topological surface Fermi-arc states.[1,20-23] In reality, the tunable carrier density plays a key role in investigating these novel transport properties in 3D Dirac/Weyl semimetals. Thus, controllable tuning of the carrier density in 3D Dirac semimetals becomes very crucial.

Conventional gating technologies, such as the back gate[24-27] and the electrolyte gate,[28-31] could gain an effective modulation of carrier density for thin samples with very low carrier densities. However, 3D Dirac semimetal is hard to be effectively tuned by these conventional gating technologies. First, 3D Dirac semimetals usually have a conducting bulk with high carrier densities, greatly lowering the efficiency of the gating. On the other hand, the Dirac nodes of $Cd_3As_2$ could only be well maintained in the nanoplates thicker than 60 nm,[1,32] below which the Dirac node would open a gap due to the quantum confinement effect.[1,33,34] Therefore, a feasible route to effectively tune the carrier density in Dirac semimetal $Cd_3As_2$ nanoplates turns out to be highly desirable.

In this letter, we demonstrate that the carrier density of $Cd_3As_2$ nanoplates can be efficiently tuned by the thermal cycling treatment (TCT) when the samples are repeatedly annealed at 350 K for one hour. As TCT goes on, both the carrier density and mobility exhibit an anomalous evolution. The upwardly shifted Fermi energy not only blurs the surface Fermi-arc states, but also changes the band topology, resulting in an anomalous phase shift of the bulk quantum oscillations. Meanwhile, the coupling between the quantum oscillation of the 2D Fermi-arc states and those of bulk states leads to a shift of peaks of bulk oscillations of the longitudinal magnetoresistivity at high fields.

The crystal structure of $Cd_3As_2$ can be regarded as distorted anti-fluorite structure with exactly ordered 1/4 Cd vacancies arranging helically along [001] direction in the ideal lattice.[35] In this tetragonal structure, there are 160 atoms in a unit cell with the arsenic atoms forming a closed-packed cubic and the cadmium atoms being four-coordinated by arsenic, as shown in Figure 1a. In our experiments, the $Cd_3As_2$ nanoplates were grown by chemical vapor deposition (CVD) method. The scanning electron microscopy (SEM) image in Figure 1b shows that the length of nanoplates can reach 100 $\mu m$, which is much longer than those in previous works (~10 $\mu m$).[36,37,38] The characterization of the crystal structure of $Cd_3As_2$ nanoplates by the high-resolution transmission electron microscopy (HR-TEM) is illustrated in Figure 1c. The HR-TEM and the selected area electron diffraction (SAED) pattern in the in Figure 1d demonstrate a [221] zone axis, indicating that the naturally grown surface is the (112) plane. The $Cd_3As_2$ nanoplates then were fabricated into the standard Hall-bar configurations and coated with Ti /Au electrodes (10/120 $nm$) by using standard electron-beam lithography (EBL) and lift-off techniques. We would like to choose the thick samples ($> 70\ nm$) to prevent the gap opening of the bulk states due to the quantum confinement effect and ensure the topologically semimetallic nature of Dirac fermions in $Cd_3As_2$ nanoplates. A typical Hall-bar device (sample #1) with the thickness of about $90\ nm$ is presented in the inset of Figure 1a. To avoid contamination or oxidation at the surface, the Hall-bar devices were covered with poly-methyl methacrylate layers before the transport measurements in a commercial Physical Property Measurement System with magnetic field up to 14 T. During the experiments, the processes of both the heat up and cool down are at a rate of 5 K/min, while the magnetic field is swept at a rate of 30 Oe/s at 2 K.

Figure. 2a shows the LMR of sample #1 at 2 K for the magnetic field $B$ parallel ($\theta = 90°$) and perpendicular ($\theta = 0°$) to the electric current $I$. For $\theta = 90°$, only single periodic Shubnikov-de Haas (SdH) oscillations are revealed and superimposed on the negative LMR. The negative LMR in $Cd_3As_2$ could be attributed to the chiral anomaly.[13,14,18] The geometry of our samples and devices could allow us to rule out some other mechanisms, such as the current jetting effect[39,40] and the fluctuation of conductivity.[41] For $\theta = 0°$, the large positive LMR exhibits multi-periodic oscillation patterns, which might originate from the unusual Weyl magnetic orbits connecting the Fermi-arcs on the opposite surfaces via the bulk chiral modes.[42] These multi-periodic oscillations were observed in samples #2 and #3 as well (see Supporting Information). If we heat up the sample *in-situ* to 350 K and keep it for one hour (at helium gas ambient with pressure less than 1 *Torr*) followed by cooling down to 2 K, the Hall coefficient $R_H = 1/ne$ (negative) increases correspondingly. Repeating the TCT for four times, marked "1st", 2nd", "3rd" and "4th" shown in Figure 1b, the Hall coefficient $R_H$ continues to increase, indicating an enhancement of the carrier density. This can also be seen in the inset of Figure. 2b, where the carrier density changes from $8.1 \times 10^{17}\ cm^{-3}$ (without TCT) to $1.17 \times 10^{18}\ cm^{-3}$ after the four consecutive TCTs. The carrier density estimated from the Hall measurement is well consistent with that from the SdH oscillations of bulk states (from $8.4 \times 10^{17}\ cm^{-3}$ (without TCT) to $1.5 \times 10^{18}\ cm^{-3}$ after the four consecutive TCTs). It suggests that the measured carrier density comes from the 3D bulk electrons in $Cd_3As_2$ nanoplates. For a band with Dirac dispersion, $E(\bm{k}) = \hbar v_F |\bm{k}|$ with $\hbar$ the Planck's constant and $v_F$ the Fermi velocity, the carrier density $n$ is $n = k_F^3/3\pi^2$ and the Fermi energy becomes $E_F = \hbar v_F (3\pi^2 n)^{1/3}$ with $k_F$ being the Fermi wave vector. Therefore, the increase of $n$ in sample #1 shifts $E_F$ upwardly away from the Dirac points. This upward shift of $E_F$ allows us to study the magnetotransport properties of the Fermi-arc surface states and their coupling to the bulk states. The Hall resistivity measurements in samples #2 and #3 also exhibit an evident enhancement of the carrier density after the first TCT, as shown in Figures. 2c-2d.

In order to gain more insight into the impacts of TCT on the carrier density, we have performed more measurements on samples #2 and #3, as shown in Figure. 3. It is worth pointing out that the carrier density in both samples is much lower than that in #1. The carrier density and mobility without TCT in these three samples have been presented in table S1 (see Supporting Information). As shown in Figure. 3a, the carrier density in sample #2 is sharply increased after the first TCT and then becomes saturated, forming a plateau. After the first five consecutive TCTs, the carrier density begins to decrease gradually and ultimately tends to be saturated again. Usually, the annealing largely reduces the defects in materials and strongly enhances the transport lifetime $\tau_{tr}$. According to the mobility of 3D Dirac fermions $\mu = ev_F\tau_{tr}/\hbar k_F$, the carrier mobility should increase. However, as shown in Figure. 3a, the carrier mobility also exhibits an anomalous behavior but with an opposite tendency compared to the carrier density after the TCTs. The similar behaviors of the carrier density and mobility were also observed in sample #3 depicted in Figure. 3b.

Let us turn to discuss the possible mechanisms for the anomalous evolution of the carrier density and the mobility induced by the TCT in $Cd_3As_2$ nanoplates. Previously, the instability of resistivity in $Cd_3As_2$ single crystals at room temperature had been reported, and some mechanisms such as surface oxidation or degradation are suggested.[8] In fact, the surface oxidation is unlikely in our samples since the surfaces are coated with poly-methyl methacrylate layers during the whole transport measurements and the TCTs are carried out *in-situ* in helium gas ambient. The structural phase transition induced by TCT can be also ruled out as the lowest transition temperature in $Cd_3As_2$ (about 220 °C)[43] is much higher than the maximal temperature in our experiments (~75 °C). A possible mechanism may be related to the annealing, which affects the cadmium vacancies[44,45] and develops the charge puddles,[46] leading to a dramatic fluctuation of the carrier density in $Cd_3As_2$ crystals.

However, the carrier densities of crystals are consecutively decreased after the annealing at room temperature in previous experiments,[44,45,46] which is strongly contrast with our results of nanoplates, as shown in Figure 3. At present, the physical mechanism of TCT induced anomalous evolution of the carrier density and the mobility is still lacking, but it certainly deserves a thorough theoretical study.

As mentioned above, without the TCT, the LMR of sample #1 exhibits multi-periodic oscillation patterns in the perpendicular field direction ($\theta = 0°$) at 2 K. Tilting the field away from the direction $\theta = 0°$ makes the multi-periodic oscillation patterns disappear gradually, as shown in Figure 4a. Above $\theta = 56°$, only single oscillation patterns are revealed and the oscillation frequency keeps unchanged (as indicated by the grey arrows), revealing an isotropic Fermi surface in $Cd_3As_2$ nanoplates. According to our previous study,[47] the extra oscillation patterns near $\theta = 0°$ at high field region might originate from the 2D oscillations of Fermi-arc states on the opposite surfaces mediated by the bulk chiral modes in the zero Landau levels. This 2D surface oscillation patterns can be testified by the angular-dependent LMR oscillations. In Figure 4b, the oscillation patterns with kinks or peaks (the black dashed lines) clearly exhibit a 2D character. Noted that, in Figure 4a, the LMR oscillation peaks from bulk states are shifted while below $\theta = 47°$ (deviated from the grey arrows), as marked by the red dashed curves. Generally, the quantum oscillations from the 2D surface states at high field are superimposed on the 3D bulk states without shifting the oscillation peaks from the bulk.[47] In reality, it is the first time that a shift of oscillation peaks of bulk states due to a strong coupling between the surface Fermi-arcs and the bulk states was observed.

To illustrate the evolution of the 2D Fermi-arc surface states with respect to the bulk Fermi energy, we now present the detailed oscillation patterns at $2\ \text{K}$ and $\theta = 0°$ both without and with the TCTs of sample #1 in Figure 4c. Firstly, the testified surface state oscillations become blurred after the TCTs, revealing a weak coupling between the surface to the bulk states. Secondly, the peak spacing of 2D surface state oscillations gets narrow, demonstrating that the oscillation frequency of surface Fermi-arcs increases as the upward shifting of the Fermi energy, as marked by the red arrows in Figure 4c. Theoretically, the oscillation frequency of the Weyl magnetic orbits $F_s$ is proportional to the Fermi energy $E_F$, $F_s = E_F k_0/(e\pi v_F)$,[42] where $k_0$ is the length of the Fermi-arc states $(\sim 0.1\ \text{Å}^{-1})$.[8,42] $k_0/(k_F^2 L)$ is the ratio of the surface to the bulk states, where $L$ is the thickness of the Dirac semimetal nanoplate between two opposite surfaces. As the Fermi energy shifts upwardly relative to the Dirac nodes, the effective length of the Fermi-arc state becomes shorten and the ratio of the surface to the bulk states decreases such that $F_s$ increases. One can see that our observations can be qualitatively consistent with the previous theoretical work.

The quantum oscillations at different Fermi energies can help us to identify the energy band and the topology of 3D Dirac semimetal. Figure 5a shows the oscillation components of LMR $\Delta\rho_{xx}$ versus $1/B$ with parallel electromagnetic fields $(\theta = 90°)$ for each measurement without and with the TCT in sample #1 at $T = 2\ \text{K}$. We present the corresponding Fast Fourier transformation (FFT) spectra for each measurement to obtain the Fermi wave vector $k_F$ in Figure 5b. As we can see, the oscillation frequency increases from about 30 T (without TCT) to 40 T after four consecutive TCTs, as indicated by the red arrows. According to the Onsager relationship, $F(T) = \frac{\hbar}{2\pi e} A_F$ with $A_F = \pi k_F^2$ being the maximal cross-sectional area of Fermi surface, we find that the magnitude of $k_F$ changes from $3.0 \times 10^{-2} \text{Å}^{-1}$ to $3.5 \times 10^{-2} \text{Å}^{-1}$ correspondingly. We extract the phase factor $\phi$ through the Landau level (LL) fan diagrams for each measurement in sample #1, in which we assign the oscillation peaks to the integer LL indices (Figure 5c).[48] The phase factor deduced from the Lifshitz-Onsager relation $\frac{\hbar}{eB} A_F = 2\pi(n + \phi)$,[49,50] is shown in the inset of Figure 5c. As we can see, the phase factor in our experiments changes monotonously from $-0.18$ to $-0.50$, approaching to $-5/8$ for 3D electron gases with a parabolic dispersion. Such a phase shift reveals a clear change of topology of the energy band from a topologically nontrivial band (linear dispersion) to a trivial band structure (parabolic dispersion) as the Fermi energy shifts upwardly relative to the Dirac nodes. The behavior of quantum oscillations with respect to the Fermi energy can be qualitatively understood through the effective model from the *ab initio* calculations.[1,48] In addition, after the fourth TCTs, the observed negative LMR is largely suppressed, while the Fermi energy might be close to the Lifshitz transition point (the detailed discussion is presented in the Supporting Information). Thus, our results provide new insights into the understanding of the energy band and its topology of 3D Dirac semimetal $Cd_3As_2$ nanoplates.

In summary, we have found that the thermal cycling treatments can effectively tune the electron density of $Cd_3As_2$ nanoplates, leading to an anomalous evolution of both the carrier density and the mobility. The upward shift of the bulk Fermi level suppresses the surface Fermi-arc states and accompanies with an anomalous phase shift of the bulk quantum oscillations. Moreover, the quantum oscillations stemming from the 2D surface Fermi-arcs are observed and make the oscillation peaks of the longitudinal magnetoresistivity of bulk states shift. The possible origins of the anomalous evolution of the carrier density and mobility are discussed. Therefore, our work provides a feasible way to effectively manipulate the exotic quantum states through the carrier density in Dirac semimetal $Cd_3As_2$ nanoplates and facilitates their applications in nanoelectronics technology.

## Acknowledgments

This work was supported by the National Key Research and Development Program of China No.2016YFA0401003, and the Natural Science Foundation of China (Grant No.11774353, No.11574320, No.11374302, and Nos.U1432251, U1732274), and the CAS/SAFEA international partnership program for creative research teams of China, and the Innovative Program of Development Foundation of Hefei Center for Physical Science and Technology (2018CXFX002), and the 100 Talents Program of Chinese Academy of Sciences (CAS). G. Z. and L.W. were supported by the Australia Research Council Centre of Excellence in Future Low-Energy Electronics Technologies (Project no. CE170100039). G.Z. was also supported by the Natural Science Foundation of China (Grant No.11804340) and the Project funded by China Postdoctoral Science Foundation (Grant No. 2018M630718).

# Figure Captions

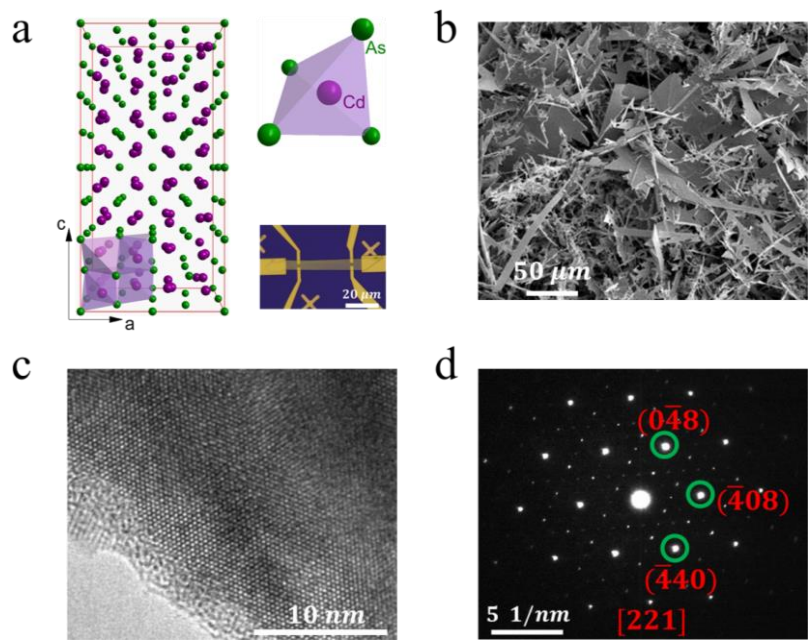

**Figure 1.** The crystal structure of Cd$_3$As$_2$. (a) Shows the crystal structure of Cd$_3$As$_2$ with a body centered tetragonal structure with $C_{4v}^{12}$ symmetry. Upper inset: the basic structure unit of Cd$_3$As$_2$. Lower inset: the optical microscope image of sample #1 with the thickness of about $90\ nm$. (b) The SEM image of CVD-grown Cd$_3$As$_2$ nonoplates. Scale bar: $50\ \mu m$. (c) The high-resolution transmission electron microscopy image of Cd$_3$As$_2$ nanoplate. Scale bar: $10\ nm$. (d) The corresponding selected area electron diffraction pattern.

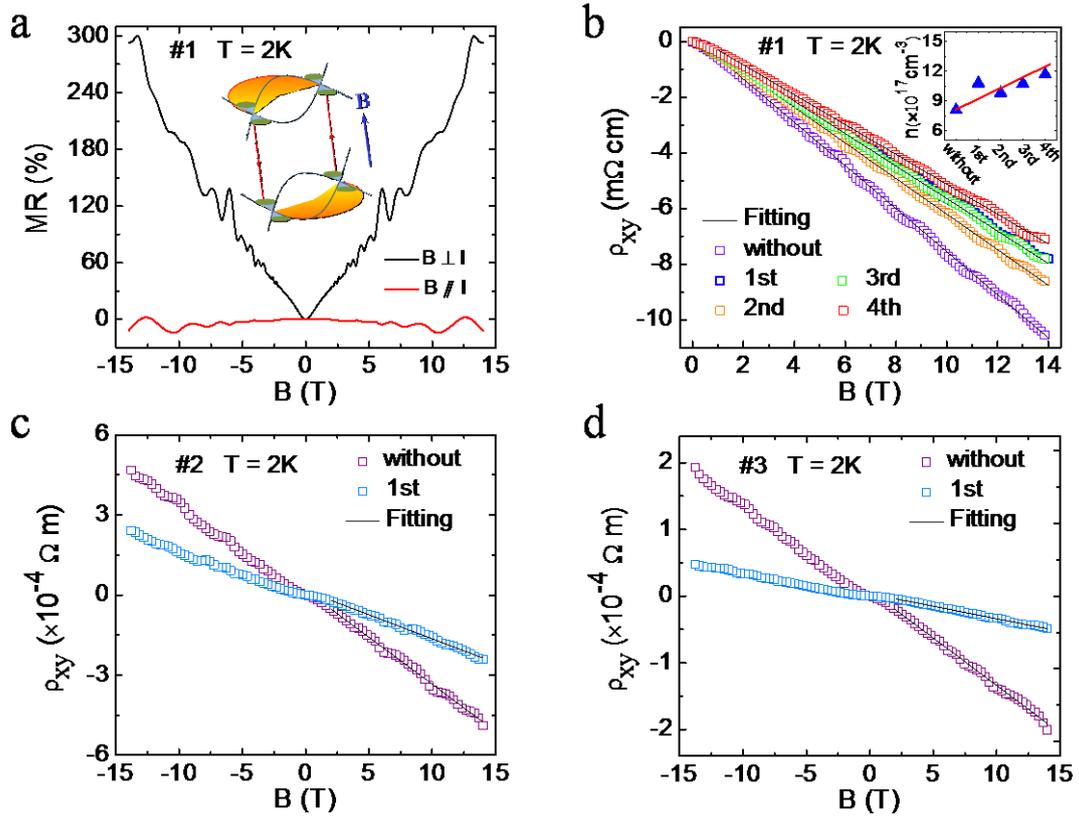

**Figure 2.** Magnetotransport properties of $Cd_3As_2$ nanoplates. (a) Multi-periodic (black) and single periodic (red) Shubnikov-de Haas oscillations observed in sample #1 at $T = 2\ K$ with field $B$ oriented perpendicular and parallel to the electric current, respectively. Inset: an illustration of Weyl magnetic orbits connecting the Fermi-arcs on opposite surfaces via the bulk chiral modes. (b) The Hall resistivity without (marked "without") and with (marked "1st", "2nd", "3rd" and "4th") the TCTs in sample #1. Inset: the corresponding carrier density for each measurement. (c-d) The Hall resistivity without and with 1st TCT in samples #2 and #3 with the thickness of about $100\ nm$ and $120\ nm$, respectively.

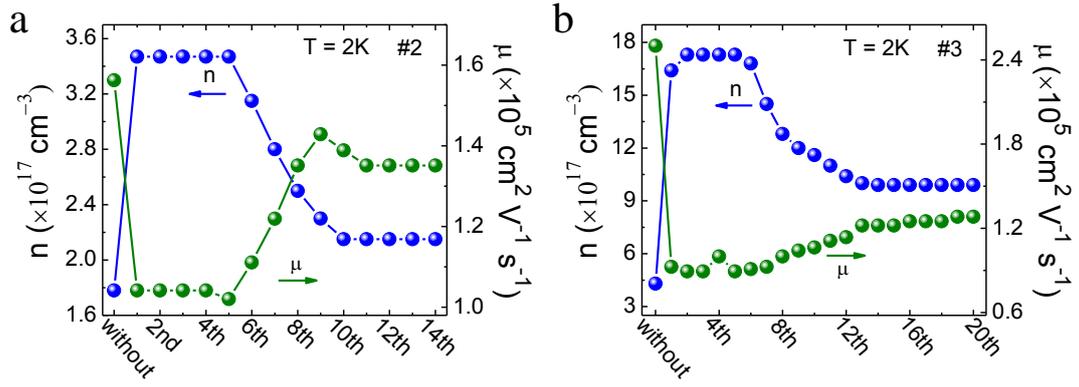

**Figure. 3** The anomalous evolution of the carrier density and mobility. (a) The TCT induced evolution of carrier density and mobility in sample #2. These two effects exhibit opposite tendency with the TCTs. Such similar behavior is also revealed in sample #3 in (b).

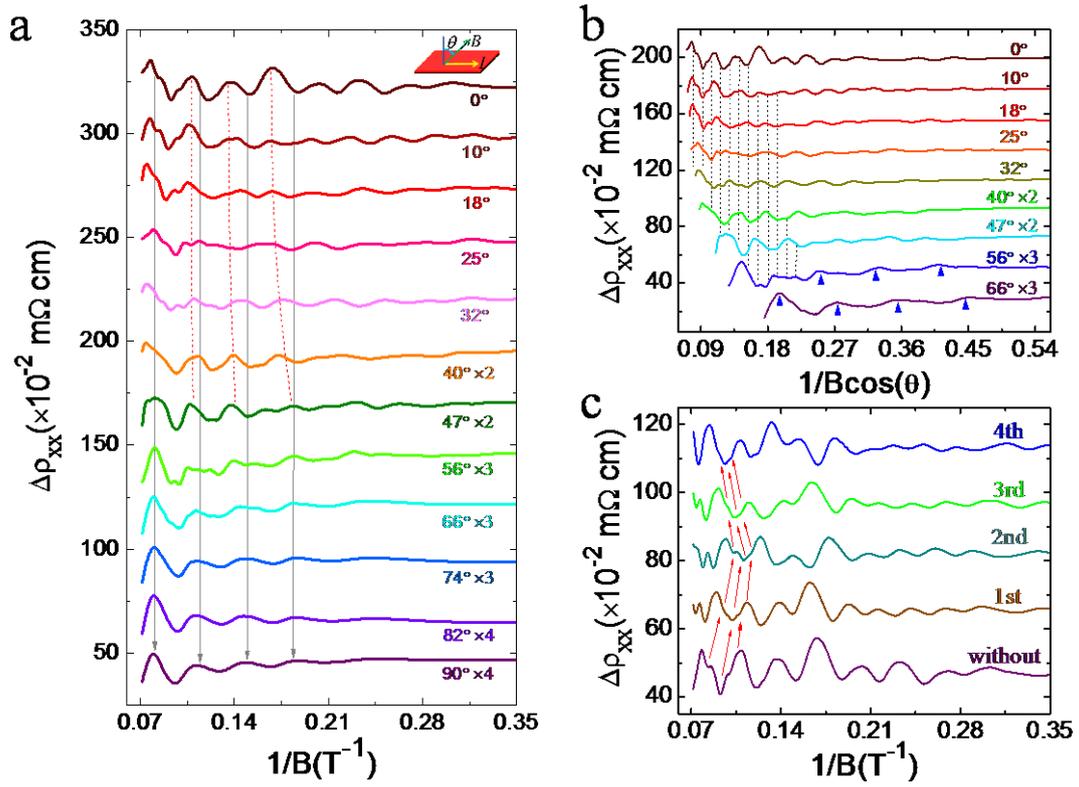

**Figure. 4** The evolution of surface Fermi-arc states in sample #1. (a) The angular-dependence of oscillation components of longitudinal magnetoresistivity $\Delta\rho_{xx}$ without TCT at $T = 2$ K. (b) The oscillation components $\Delta\rho_{xx}$ versus $1/B\cos\theta$ at various titled angles without TCT. Both 2D oscillations from the surface Fermi-arcs (black dashed lines) and 3D bulk oscillations (blue arrows) are revealed. (c) Plots the $\Delta\rho_{xx}$ versus $1/B$ for each measurement without and with the TCTs at $T = 2$ K, $\theta = 0°$. The red arrows indicate the evolution of surface Fermi-arc states.

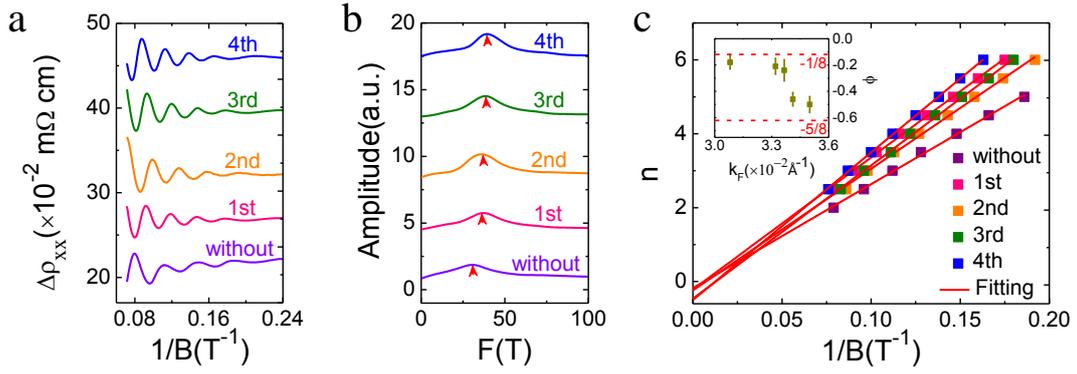

**Figure 5** The anomalous phase shift of bulk quantum oscillations in sample #1. (a) The oscillation amplitudes $\Delta\rho_{xx}$ versus $1/B$ for each measurement without and with the TCTs in the presence of parallel magnetic and electric fields at $T = 2$ K. (b) The corresponding FFT spectra. (c) The Landau level index $n$ versus $1/B$ for each measurement. Inset: shows the phase factor changes from $-0.18$ to $-0.5$ as the increase of the Fermi wave vectors.